# Generalized Integral Siding Mode Manifold Design: A Sum of Squares Approach

S.Sanjari, *Member, IEEE*, S.Ozgoli, *Member, IEEE*

*Abstract*— this paper presents a general form of integral sliding mode manifold, and proposes an algorithmic approach based on Sum of Squares (SOS) programming to design generalized integral sliding mode manifold and controller for nonlinear systems with both matched and unmatched uncertainties. The approach also gives a sufficient condition for successful design of controller and manifold parameters. The result of the paper is then verified by several simulation examples and two practical applications, namely Glucose-insulin regulation problem and the unicycle dynamics steering problem are considered.

## I. INTRODUCTION

Sliding mode control (SMC) is one of the most effective control methodologies in dealing with a large class of uncertain systems. The controller consists of a high-frequency switching term that completely compensates matched perturbations (i.e. perturbations acting in the direction of control input). This action takes place when state trajectory remains on the subspace of the state space called "sliding manifold". Definition of a suitable sliding manifold is nevertheless an open problem in SMC theory. The difficulty arises when answering two questions: what features does each manifold possess? and how the parameters of these manifolds and controllers can be found?

In this regard, one choice for sliding manifold is the integral surface first proposed in [1] and developed for unmatched perturbation in [2] . The main feature of Integral Sliding Mode Controller (ISMC) is the elimination of reaching phase achieved by proper sliding manifold design [1]. Compensated system in this type of SMC has full order (i.e. the order of the closed loop system is equal to that of the original uncompensated system when sliding motion takes place). The integral sliding manifold and ISMC are designed to completely reject uncertainties and make the closed loop system act in the same manner as the nominal system.

The linear sliding manifold presented in [3] minimizes the influence of unmatched perturbation on closed loop dynamics for linear systems and for linear manifold. A nonlinear extension of this manifold is given in [4] for a class of nonlinear systems with unmatched perturbations that satisfy involutive condition. The drawback of this method is the difficulty in obtaining manifold and controller parameters which calls for a systematic approach to be developed. To overcome the problem, the Linear Matrix Inequality (LMI)-based method of [5] works well for linear systems and manifolds, however, for the nonlinear case, no systematic method has been introduced yet. Addressing this problem is the main objective of the present study.

In this article, we first introduce the generalized integral sliding mode manifold, and then give an algorithmic design procedure for nonlinear systems based on SOS programming [6, 7]. Next, a special case of this systematic approach, solvable by linear semi definite program, is proposed for nonlinear integral manifold given by [4]. The integral sliding manifold proposed in this article is a generalization of integral sliding mode manifold proposed in [4]; therefore, sliding manifold proposed in [4] is only a special case of generalized integral sliding which can be obtained constructively. On the other hand, to simplify the computational approach, a constructive algorithm based on SOS is also proposed to obtain parameters of control and manifold of ISMC introduced in [4].

The SOS technique is a cornerstone of the algorithm developed in this paper. This technique is originally used for systems with polynomial or rational vector fields, but thanks to its extension to non-polynomial systems [8, 9], its applicability has also been significantly extended. It in fact provides a proper relaxation for control problems by using SOS decomposition and semi-definite programming (SDP) for nonlinear systems. For example, an algorithmic method based on SOS technique has been proposed to generate Lyapunov function [8] and also to design state feedback controller [10]; and ultimately, absolute stability [9], and estimation of region of attraction [11] have been examined by this method illustrating the effectiveness of SOS programming in control problems.

The rest of the paper is organized as follows: a brief review On SOS approach and preliminaries are presented in section II. Section III discusses the mentioned control problem and presents the general dynamics of target systems. SOS-based programming, used for designing the ISMC, is then presented in section IV. In section V, unmatched perturbation has been added to system, and a combination of ISM and $H_\infty$ controller

S. Sanjari is with Tarbiat Modares University, Tehran, IRAN. (e-mail: s.sanjari@modares.ac.ir).
S. Ozgoli is with Tarbiat Modares University, Tehran, IRAN. (e-mail: ozgoli@modares.ac.ir).

has been designed using SOS technique. In section VI, three numerical examples are given to show the effectiveness of the presented method. Applications of the proposed method in glucose-insulin regulatory system of human body and unicycle dynamics are given in section VII. Finally, section VIII concludes the paper.

## II. Preliminaries

This section presents a brief review on SOS decomposition, and other definitions needed to follow the paper.

*Definition 1 (Monomial)*: A monomial $Z_\alpha$ is a function defined as

$$Z_\alpha = x_1^{\alpha_{i_1}} x_2^{\alpha_{i_2}} \ldots x_n^{\alpha_{i_n}}$$

For $\{\alpha_{i1}, \ldots, \alpha_{in}\} \in Z_+$, and its degree is given by $\deg(Z_\alpha) = \sum_{i=1}^{n} \alpha_i$.

*Definition 2 (Polynomial)*: a real polynomial function $p \in R[x] = \mathcal{R}[x_1 \ldots, x_n]$ is defined as

$$p(x) = \sum_k C_k Z_{\alpha_k}$$

where $C_k \in \mathbb{R}$ and $x \in \mathbb{R}^n$. The polynomial $p(x)$ is said to be of degree $m$ if it corresponds to the largest monomial degree in $p(x)$ i.e. $m = max_k \deg(Z_{\alpha k})$.

In most control problems, "Lyapunov problem" for example, it is important to investigate the non-negativity of polynomials. In general, it is extremely hard or sometimes even impossible to solve this problem. However, checking whether a polynomial is sum of squares or not is a SDP which can be easily done. So, in our problem formulation, conditions on non-negativity are replaced by sufficient conditions for polynomials to be SOS.

*Definition 3 (SOS)*: a real polynomial $p(x) \in \mathcal{R}_n$ of degree $d$ is SOS if there exist polynomials such that

$$p(x) = \sum_{i=1}^{r} p_i^2(x)$$

Additionally, the subset of all SOS polynomials in $\mathcal{R}_n$ is denoted by $\Sigma_n$.

The SOS definition implies that the existence of SOS decomposition is sufficient condition for $p(x)$ to be positive semidefinite, i.e. $p(x) \geq 0$. In general, the converse of this result does not hold; however, the possibility of $\mathcal{R}_n$ being $\Sigma_n$ has been calculated in [16]. It is demonstrated that the gap between these two set is negligible.

*Lemma 1 (S-procedure) [8]*: given $\{p_i\}_{i=0}^{m} \epsilon \mathcal{R}_n$, if there exist $\{s_i\}_{i=1}^{m} \epsilon \Sigma_n$ such that $p_0 - \sum_{i=1}^{m} s_i p_i \epsilon \Sigma_n$, then $\cap_{i=1}^{m} \{x \in \mathbb{R}^n | p_i(x) \geq 0\} \subseteq \{x \in \mathbb{R}^n | p_0(x) \geq 0\}$.

*Lemma 2:[10]* for a symmetric polynomial matrix $p(x)$ that is non-negative for all $x$, the following equality holds.

$$\frac{\partial p}{\partial x_i}(x) = -p(x) \frac{\partial p^{-1}(x)}{\partial x_i} p(x)$$

*Notation:* for matrix $Q \in \mathbb{R}^{n \times n}$, $Q \geq 0$ represents positive semi-definiteness of $Q$; $Q(x) \in \mathcal{R}[x]$ means that $Q(x)$ is a polynomial; $Q(x) \in \mathcal{R}^c[x]$ means that $Q(x)$ is a c-dimensional polynomial vector; $Q^+(x)$ is a left pseudo-inverse of $Q(x)$, i.e. $Q^+(x) \triangleq (Q^T(x)Q(x))^{-1}Q^T(x)$. $\|a\|$ denotes the 2 norm of $a$.

## III. System Description and Problem Statement

Consider the following nonlinear uncertain system:

$$\dot{x} = f(x) + B(x)u(t) + B(x)\varphi_0(t,x)u(t) + \xi(t,x) \quad (1)$$

Where $x \in \mathbb{R}^n$ is the state vector, $u \in \mathbb{R}^m$ is the control input, $f(x) \in \mathbb{R}^n$ is a known nonlinear function, and $B(x) \in \mathbb{R}^{n \times m}$ is a known full rank state-dependent matrix. $\xi(x,t)$ is a function that models both matched and unmatched perturbation terms, and $\varphi_0(x,t)$ represents multiplicative uncertainty in control.

In section IV perturbation is considered to be matched which is modeled as $B(x)\varphi_1(x,t)$. Consequently the system equations can be written as:

$$\dot{x} = f(x) + B(x)\{(I + \varphi_0(x,t))u(t) + \varphi_1(x,t)\} \quad (2)$$

This assumption is not very restrictive, and is made by several relevant papers (see [12] for instance). It will however be relaxed in section V. The following model describes system with both matched and unmatched perturbations.

$$\dot{x} = f(x) + B(x)\{(I + \varphi_0(x,t))u(t) + \varphi_1(x,t)\} + B^\perp(x)\varphi_2(t,x) \quad (3)$$

Where $B^\perp(x) \in \mathbb{R}^{n \times (n-m)}$ is a known matrix spanning null space of $B(x)$.

The general model considered in this paper is (1). This model is made simple as (2) in section IV and without simplification is considered as (3) in section V. the following assumptions are made on these models.

*Assumption 1*: Although perturbations are considered to be unknown, they are assumed to be bounded i.e.

$$\|\varphi_0(x,t)\| \leq \beta_0 \quad (4)$$

Where $\beta_0 < 1$ is a positive number, and

$$\|\varphi_1(x,t)\| \leq \beta_1(x,t) \quad (5)$$
$$\|\varphi_2(x,t)\| \leq \beta_2(x,t) \quad (6)$$

∎

*Assumption 2*: The distribution $\Delta(x) = span\{B_i^\perp(x)\}$ is involutive [13] i.e.

$$[B_i^\perp(x), B_j^\perp(x)] \in \Delta(x) \quad (7)$$

Where $i,j = 1, \ldots, n - m$, and $B_i^\perp$ stands for the i-th column of $B^\perp$. $[.,.]$ is the Lie bracket of two vector fields:

$$[B_i^\perp(x), B_j^\perp(x)] = \frac{\partial B_j^\perp(x)}{\partial x} B_i^\perp(x) - \frac{\partial B_i^\perp(x)}{\partial x} B_j^\perp(x)$$

∎

In order to design a sliding mode controller, the following nonlinear integral-type sliding mode manifold is considered

$$s(x,t) = g(x) + z(t) \quad (8)$$

Where $g(x): \mathbb{R}^n \to \mathbb{R}^m$, and $z(t): \mathbb{R}^n \to \mathbb{R}^m$ are nonlinear functions and $z(t)$ is generated by another nonlinear function $D(x): \mathbb{R}^n \to \mathbb{R}^m$ as follows

$$\begin{cases} \dot{z}(t) = D(x) \\ z(x_0) = -g(x_0) \end{cases} \quad (9)$$

The initial condition in (9) is checked such that the system would be restricted to sliding manifold from the initial time instant, i.e. $s(x_0, t_0) = 0$.

The objective of this article is to systematically determine the parameters of the sliding manifold and controller so that system becomes asymptotically stable.

IV. NONLINEAR INTEGRAL SMC: MATCHED PERTURBATION

This section concentrates on stabilizing a system with only matched uncertainty which means that uncertainty is only contained in the input channel. Theorem 1 is accordingly presented to give sufficient conditions based on SOS constraints which can be translated by semi-definite program which is solvable by SOSTOOLS toolbox [14]. Afterwards, Theorem 2 characterizes a special case of Theorem 1 in which integral sliding manifold restricts closed loop dynamics to nominal dynamics (presented in [4]).

A. General Integral sliding surface

The following Theorem shows how sliding manifold and controller parameters are designed.

*Theorem 1*: The uncertain system (2) which satisfies assumptions 1 and 2 will be asymptotically stable by applying the following control law

$$u(t) = \begin{cases} -\rho(x,t) \frac{\gamma(s)}{\|\gamma(s)\|} & s(x,t) \neq 0 \\ 0 & s(x,t) = 0 \end{cases} \quad (10)$$

Where $s(x,t)$ is defined in (8), (9), and $\gamma(s)$ is chosen to be a nonlinear function with $\gamma(s) = 0$ only if $s = 0$. $\rho(x,t)$ is the switching gain function which is chosen so that satisfies the following inequality.

$$\rho(x,t) \geq \frac{1}{1+\beta_0}\left(\beta_1(x,t) + \frac{1}{\|MB\|}\|Mf + D\|\right) \quad (11)$$

Where $M(x)$ is the Jacobian matrix of $g(x)$. The unknown functions $M$, $g$ and $D$ are constructed by
1) Choosing small constants $\varepsilon_{ij}$ and constructing

$$l_k(x) = \sum_{i=1}^{n}\sum_{j=1}^{d} \varepsilon_{ij} x_i^{2j}, \sum_{j=1}^{m} \varepsilon_{ij} > 0, \forall i,j = 1,\ldots,n, \varepsilon_{ij} \geq 0, k = 1,2$$

2) Solving the following SOS program

Find polynomial $V(x), V(0) = 0$ and $m \times 1$ polynomial vectors $K(x)$, $D(x)$ and $g(x)$, and a $m \times m$ positive polynomial matrix $L(x)$

$$V - l_1 \in \Sigma_n \quad (12)$$
$$-\frac{\partial V}{\partial x}\{(I - BB^+)f(x) - B(B^T B)^{-1}K\} - l_2 \in \Sigma_n \quad (13)$$

Whit constraints

$$LK = D \quad (14)$$
$$\frac{\partial g}{\partial x} = LB^T \quad (15)$$

**Proof**: In order to show the asymptotic stability of sliding mode dynamics, we first prove that the control law guarantees sliding mode behavior. Second, we derive the sliding mode dynamics using the equivalent control method [15] and finally, we prove that conditions for asymptotic stability of the sliding mode dynamics based on Lyapunov approach can be satisfied by the sum of squares program of the theorem.

To prove that the above controller can maintain the sliding mode, we show that reaching condition is satisfied.

$$s^T \dot{s} = s^T M(x)[f(x) + B(x)u(t) + B(x)\varphi_0(x,t)u(t) + B(x)\varphi_1(x,t)] + s^T D(x)$$

$$= s^T\left\{(MB)(I + \varphi_0(x,t))\left(-\rho(x,t)\frac{\gamma(s)}{\|\gamma(s)\|}\right) + \varphi_1(x,t)\right\}$$
$$+ s^T\{M(x)f(x) + D(x)\}$$
$$\leq -\|s\|\|MB\|\{(1 + \beta_0)\rho(x,t) - \beta_1(x,t) - \frac{1}{\|MB\|}\|Mf + D\|\}$$

So the reaching condition is satisfied which ensures finite time stability [12]; therefore, switching gain function satisfying (11) guarantees that the sliding mode can be maintained, $\forall t \in [t_0, \infty)$.

Set $s = \dot{s} = 0$. The equivalent control law is now obtained as

$$u_{eq} = -(I + \varphi_0(x,t))^{-1}(M(x)B(x))^{-1}(M(x)f(x) + D(x) + M(x)B(x)\varphi_1(x,t)) \quad (16)$$

Substituting equivalent control (16) into (2), one obtains sliding mode dynamics:

$$\dot{x} = \left(I - B(x)(M(x)B(x))^{-1}M(x)\right)f(x) - B(x)(M(x)B(x))^{-1}D(x) \quad (17)$$

Now consider function $V$, the output of the above SOS program as a lyapunov candidate function. Due to (12), $V$ is positive definite function. Calculate its time derivate:

$$\dot{V} = \frac{\partial V}{\partial x}\left\{\left(I - B(x)(M(x)B(x))^{-1}M(x)\right)f(x) - \right.$$

$B(x)(M(x)B(x))^{-1}D(x)\}$

Where $(M(x)B(x))$ is a full rank matrix.

Assumption 2 is sufficient condition to the existence of $g, L$ such that (15) is satisfied [4], So $\dot{V}$ can be written as

$$\dot{V} = -\frac{\partial V}{\partial x}\{\left(I - B(x)(M(x)B(x))^{-1}M(x)\right)f(x) \\ - B(x)(M(x)B(x))^{-1}D(x)\} \\ = -\frac{\partial V}{\partial x}\{(I - BB^+)f(x) \\ - B(B^TB)^{-1}L^{-1}D\}$$

Now (13) implies that $\dot{V}$ is negative definite, so $V$ is a lyapunov functions and the proof is concluded. ∎

*Remark 1:* Note that assumption (2) is not needed to be satisfied in the SISO case and (13) in SOS program can be replace by

$$-\frac{\partial V}{\partial x}\{(M(x)B(x)I - B(x)M(x)))f(x) - B(x)D(x)\} - l_2 \in \Sigma_n \tag{18}$$

*Remark 2:* With regard to the definition of $f(x)$, this function can contains non-polynomial terms. However, SOS approach is presented solely for polynomial vector fields. In order to handle this problem, one way is to consider all non-polynomial terms as perturbation. On the other hand, this may lead to increase in the bounds of perturbation. In addition, this idea is not applicable to some cases since the main part of system may consist of non-polynomial terms such as the case in study B (unicycle application). In order to solve this problem, we can use the recasting procedure (see [8]) or functional approach (see [9]) to transform non-polynomial system into a polynomial one. In recasting procedure, non-polynomial system, which consists of elementary function, is converted to polynomial system by defining slack variables. Thus, constraint (12) and (13) are restated respectively as follows:

$$V - l_1(\bar{x}_1, \bar{x}_2) - \lambda_1^T(\bar{x}_1, \bar{x}_2)G_1(\bar{x}_1, \bar{x}_2) - \\ \sigma_1^T(\bar{x}_1, \bar{x}_2)G_2(\bar{x}_1, \bar{x}_2) \in \Sigma_n \tag{19}$$

$$-\frac{\partial V}{\partial x}\{(I - BB^+)f(x) - B(B^TB)^{-1}K\} - \\ \lambda_2^T(\bar{x}_1, \bar{x}_2)G_1(\bar{x}_1, \bar{x}_2) - \sigma_2^T(\bar{x}_1, \bar{x}_2)G_2(\bar{x}_1, \bar{x}_2)) \in \Sigma_n \tag{20}$$

$\bar{x}_1$ and $\bar{x}_2$ include original and slack variables of system respectively. In (19) and (20), polynomial column vectors $\lambda_1(\bar{x}_1, \bar{x}_2)$ and $\lambda_2(\bar{x}_1, \bar{x}_2)$ and sum of squares polynomial vectors $\sigma_1(\bar{x}_1, \bar{x}_2)$ and $\sigma_2(\bar{x}_1, \bar{x}_2)$ are of appropriate dimensions. $\bar{x}_1$ and $\bar{x}_2$ are such that the following constraints hold.

$$G_1(\bar{x}_1, \bar{x}_2) = 0 \tag{21}$$
$$G_2(\bar{x}_1, \bar{x}_2) \geq 0 \tag{22}$$

*Remark 3:* constraint (13) contains products of decision variable, and hence, the problem cannot be transformed into linear semi-definite program, but it can be converted to a bilinear semi-definite program solvable by PENBMI solver, a local bilinear matrix inequality (BMI) solver from PENOPT [16], or iterative method [17] or, density function [18, 19]. In order to simplify the computation of SOS program and utilizing SOSTOOLS solely to solve linear semi-definite program, sliding manifold is restricted and theorem 2 in section B is proposed.

### B. Nominal integral sliding surface

In this section, we focus on the task of finding a simple algorithm formulated in a linear semi-definite program to determine the parameters of sliding manifold and controller, when the sliding manifold is restricted to the precise function proposed by [4].

*Assumption 3*: The nominal (unperturbed) system is asymptotically stable under state feedback $k(x)$. By Lyapunov theorem, this means that there exists a nonempty set of Lyapunov functions $\mathcal{V} \in C^1$ such that for any choice of function $V(x) \in \mathcal{V}: \mathbb{R}^n \rightarrow \mathbb{R}^+$,

$$\frac{\partial V}{\partial x}[f(x) + B(x)k(x)] < 0 \tag{23}$$

∎

*Theorem 2*: The uncertain system (3) which satisfies assumptions (1) and (3) will be asymptotically stable by applying the following control law

$$u(t) = \begin{cases} q(\tilde{x})N(x)Q^{-1}(\tilde{x})Z(x) - \rho(x,t)\frac{(MB)^Ts}{\|(MB)^Ts\|} & s(x,t) \neq 0 \\ q(\tilde{x})N(x)Q^{-1}(\tilde{x})Z(x) & s(x,t) = 0 \end{cases} \tag{24}$$

Where $Z(x)$ is an $N \times 1$ vector of monomials with argument $x$ satisfying the assumption $Z(x) = 0$ if $x = 0$. Siding manifold is defined by

$$s(x,t) = g(x(t)) - g(x(t_0)) - \int_{t_0}^{t} M(x)(f(x) + B(x)q(\tilde{x})N(x)Q^{-1}(\tilde{x})Z(x))d\tau \tag{25}$$

and the switching gain function satisfies

$$\rho(x,t) > \frac{1}{1-\beta_0}(\beta_0\|k(x)\| + \beta_1(x,t)) \tag{26}$$

$N(x), q(x)$ and $Q(x)$ are found by solving the following sum of squares program:

Find polynomial matrices $N(x), Q(\tilde{x})$ and SOS polynomials $\varepsilon_2(x), q(\tilde{x})$ and positive scalar $\varepsilon_1$ such that the following two expressions are sum of squares

$$(Q(\tilde{x}) - \varepsilon_1 I) \tag{27}$$
$$-\left(q(\tilde{x})[Q(\tilde{x})A^T(x)G^T(x) + G(x)A(x)Q(\tilde{x}) + \right.$$

$$N^T(x)B^T(x)G^T(x) + G(x)B(x)N(x)] - \sum_{j \in J}[(q(\tilde{x})\frac{\partial Q(\tilde{x})}{\partial x_j} - Q(\tilde{x})\frac{\partial q(\tilde{x})}{\partial x_j})(A_j(x)Z(x))] + \varepsilon_2(x)I) \quad (28)$$

in which $Q(\tilde{x})$ and $N(x)$ are $N \times N$ symmetric and $m \times N$ polynomial matrices respectively.

**Proof**: similar to the proof of theorem 1, it can be proved that the gain function satisfying (26) guarantees that the sliding mode $(s = 0)$ can be maintained. By using the equivalent control method and setting $s = \dot{s} = 0$, equivalent control is obtained as follows:

$$u_{eq} = (I + \varphi_0(x,t))^{-1}(k(x) - \varphi_1(x,t)) \quad (29)$$

This yields closed loop dynamics as

$$\dot{x} = f(x) + B(x)k(x) \quad (30)$$

Where $k(x) = q(\tilde{x})N(x)Q^{-1}(\tilde{x})Z(x)$.
Now SOS programming is used to design the ISMC.
Consider the closed loop system as the following linear-like model.

$$\dot{x} = A(x)Z(x) + B(x)k(x) \quad (31)$$

Where $A(x)$ and $B(x)$ are polynomial matrices and $Z(x)$ is a $N \times 1$ vector of monomials with argument $x$ and $Z(0) = 0$.
Suppose that $G(x)$ is the Jacobian matrix of $Z(x)$, i.e.

$$G_{ij}(x) = \frac{\partial Z_i}{\partial x_j}(x) \quad (32)$$

for $i = 1, \ldots, N, j = 1, \ldots, n$. Let $j$ denote the rows of $B(x).J = \{j_1, \ldots, j_m\}$ shows the row indices of $B(x)$ which are equal to zero, and define $\tilde{x} = (x_{j_1}, \ldots, x_{j_m})$ in order to ensure the convexity of problem.
Define the Lyapunov function candidate for the linear-like closed loop system (32) as follows

$$V(x) = Z^T(x)p^{-1}(\tilde{x})Z(x) \quad (33)$$

Where $p(\tilde{x}) = q^{-1}(\tilde{x})Q(\tilde{x})$ is the same as $q(\tilde{x})$ in SOS polynomials. If the condition (29) and assumptions (1) and (3) are satisfied, it can be concluded that $p(\tilde{x})$ is positive definite and therefore $V > 0$ for all $x \neq 0$.

Taking derivative of the Lyapunov function with respect to time and substituting the closed loop system equations give

$$\dot{V} = Z^T(x)\left\{ (A(x) + B(x)N(x)Q^{-1}(\tilde{x}))^T G^T(x)p^{-1}(\tilde{x}) \right.$$
$$+ p^{-1}(\tilde{x})G(x)(A(x) + B(x)N(x)Q^{-1}(\tilde{x}))$$
$$\left. + \sum_{j \in J}(\frac{\partial p^{-1}(\tilde{x})}{\partial x_j}A_j Z(x)) \right\} Z(x)$$

Pre- and post-multiply both sides of the above equation by $q(\tilde{x})^2 p(\tilde{x})$ and use lemma 2, to conclude that if (29) holds with $\varepsilon_2(x) > 0$ for all $x \neq 0$, then $\dot{V}$ is negative definite and the closed loop system is asymptotically stable. Since gain function satisfies requirement of theorem 1 and closed loop dynamics is stable, controller (24) stabilizes the system represented by (2). ∎

*Remark 4*: The stability holds globally only if $p(\tilde{x})$ is a constant matrix.

In this paper, the polynomial matrix $p(\tilde{x})$ has extended the theorem provided by [10] to rational matrix case by embedding $q(\tilde{x})$, and has relaxed some assumptions. Thus a more flexible feedback control synthesis scheme has been achieved compared to [10].

V. CONSIDERING UNMATCHED PERTURBATION

This section deals with systems with both matched and unmatched perturbation. Similar to the previous section, we first present an approach to determine sliding controller and manifold in general case. Then, in order to simplify computation of approach, we also propose a constructive approach to find parameters of the sliding manifold presented in [4].

*A. Generalized manifold*

In this subsection a combination of generalized ISMC with performance constraint $H_\infty$ is designed in order to stabilize the system with both matched and unmatched perturbations using the SOS technique.

*Theorem 3*: The uncertain system (3) that satisfies assumptions 1 and 2 will be asymptotically stable by applying the controller

$$u(t) = \begin{cases} -\rho(x,t)\frac{\gamma(s)}{\|\gamma(s)\|} & s(x,t) \neq 0 \\ 0 & s(x,t) = 0 \end{cases} \quad (34)$$

Where $s(x,t)$ is defined by (8,9), $\gamma(s)$ is chosen to be a nonlinear function with $\gamma(s) = 0$ only if $s = 0$, and switching gain function satisfies

$$\rho(x,t) \geq$$
$$\frac{1}{1+\beta_0}(\beta_1(x,t) + \frac{1}{\|MB\|}(\|Mf + D\| + \|MB^\perp\|\beta_2(x,t)) \quad (35)$$

The unknown functions $M, g$ and $D$ are constructed by
1) *Choosing small constants $\varepsilon_{ij}$ and constructing*

$$l_k(x) = \sum_{i=1}^{n}\sum_{j=1}^{d}\varepsilon_{ij}x_i^{2j}, \sum_{j=1}^{m}\varepsilon_{ij} > 0, \forall i,j = 1,\ldots,n, \varepsilon_{ij} \geq 0, k = 1,2$$

2) *Solving the following SOS program*

Find $V \in \mathcal{R}_n$, $V(0) = 0$, and $m \times 1$ polynomial vectors $K(x)$, $D(x)$ and $g(x)$, and a $m \times m$ positive polynomial matrix $L(x)$

$$V - l_1 \in \Sigma_n \quad (36)$$
$$\frac{\partial V}{\partial x}\{(I - BB^+)f(x) - B(B^TB)^{-1}K\} + z^Tz - \gamma^2 w^T w - l_2 \in \Sigma_n \quad (37)$$
With constraints (14), and (15).

**Proof:** similar to the proof of theorem (1), it can be shown that gain function satisfying (35) guarantees that the sliding mode can be maintained. Equivalent control effort is given by

$$u_{eq} = -(I + \varphi_0(x,t))^{-1}\{(M(x)B(x))^{-1}(M(x)f(x) + M(x)B^\perp(x)\varphi_2(x,t) + D(x)) + \varphi_1(x,t)\} \quad (38)$$

Substituting equivalent control (38) into (3) one obtains sliding mode dynamics as

$$\dot{x} = \left(I - B(x)(M(x)B(x))^{-1}M(x)\right)f(x) - B(x)(M(x)B(x))^{-1}D(x) + \left(I - B(x)(M(x)B(x))^{-1}M(x)\right)B^\perp(x)\varphi_2 \quad (39)$$

As seen, the matched perturbation is completely compensated, but the unmatched perturbation has only transformed into a new form:

$$\varphi_{eq}(x,t) = (I - B(x)(M(x)B(x))^{-1}M(x))B^\perp \varphi_2(t,x) \quad (40)$$

Like proof of theorem (1), we assume $M(x) = L(x)B^T(x)$. It can be verified that this selection introduces a solution for the following optimization problem (see [4]).

$$\frac{\partial g^*(x)}{\partial x} \triangleq M^*(x) = argmin_{M(x)\in\mathbb{R}^{m\times n}} \|\varphi_U(x,t)\|$$

This problem has been considered in [3, 4] where it is proved that it is not possible to obtain an equivalent perturbation that has a smaller 2-norm compared to the unmatched perturbation, $\varphi_U(x,t) = B^\perp(x)\varphi_2(x,t)$. Therefore, by this selection norm 2 of the resulting equivalent disturbance (40) is equivalent to norm 2 unmatched perturbations. Taking this point into account, the sliding mode dynamics is obtained as

$$\dot{x} = (I - BB^+)f(x) - B(B^TB)^{-1}K + \varphi_{eq}(x,t)$$

In which $K(x) = L(x)^{-1}D(x)$, $\varphi_{eq}(x,t) = g_w(x)w$. We prove that SOS constraints (36) and (37) give sufficient conditions in order for the previous dynamics to be asymptotically stable, and that the induced $L_2$-gain from w to z, which is considered a performance constraint, is minimized by designing manifold parameter $K(x)$. To this end, define z as an artificial penalty variable function of state and control. Now (37) implies:

$$\frac{\partial V}{\partial x}\{(I - BB^+)f(x) - B(B^TB)^{-1}K\} + z^Tz < \gamma^2 w^T w \quad (41)$$

applying lemma 1 similar to [12], it is straightforward to show that this conditions solve the problem, so the proof is completed. ∎

*Remark 5*: in order to reduce switching gain function we can add a continuous part to control. Accordingly, define $u_0(t) \triangleq u_{01}(t) + (M(x)B(x))^{-1}u_{02}(t)$ to provide some degrees of freedom for design method. This continuous part of control can also be designed to reduce the switching gain function which leads to chattering reduction in control action. The first part can be used to attenuate the influence of matched perturbation (especially when we consider non-polynomial term as a perturbation), and the second part is used to reduce unmatched perturbation impact and sliding manifold influence on switching gain function. Similar to the proof of theorem 1 can conclude that (11) and (35) respectively change in to following inequalities

$$\rho(x,t) \geq \frac{1}{1+\beta_0}(\|u_{01}(t)\| + \beta_0\|u_0(t)\| + \beta_1(x,t) + \frac{1}{\|MB\|}\|Mf + D + u_{02}\|) \quad (42)$$

$$\rho(x,t) \geq \frac{1}{1+\beta_0}(\|u_{01}(t)\| + \beta_0\|u_0(t)\| + \beta_1(x,t) + \frac{1}{\|MB\|}\|Mf + D + u_{02} + MB^\perp\varphi_2(x,t)\|) \quad (43)$$

B. *Nominal manifold*

*Theorem 4*: The uncertain system (3) that satisfies assumption $(1 - 3)$ will be asymptotically stable by applying

$$u(t) = \begin{cases} -\gamma B_2^T G^T P^{-1}(\tilde{x})Z(x) - \rho(x,t)\frac{(L(x)B^T(x)w^{-1}(x)B)^T s}{\|(L(x)B^T(x)w^{-1}(x)B)^T s\|} & s \neq 0 \\ -\gamma B_2^T G^T P^{-1}(\tilde{x})Z(x) & s = 0 \end{cases} \quad (44)$$

The gain function $\rho(x,t)$ satisfies the following inequality.

$$\rho(x,t) > \frac{1}{1-\beta_0}(\beta_0\|k(x)\| + \beta_1(x,t) + \|B^\perp\|\beta_2(x,t)) \quad (45)$$

Sliding mode controller and manifold parameters are found by the following sum of squares program.

Find polynomials $P(\tilde{x})$ and $L(x)$, SOS polynomials $\varepsilon_2(x)$ and $w(x)$ and positive scalar $\varepsilon_1$ such that the following expressions are sum of squares.

$$(P(\tilde{x}) - \varepsilon_1 I) \quad (46)$$
$$(L(x) - \varepsilon_1 I) \quad (47)$$
$$-\begin{bmatrix} \psi_1 & PC_1^T & GB_1 \\ C_1 P & -(\gamma - \varepsilon_2)I & 0 \\ B_1^T G^T & 0 & -(\gamma - \varepsilon_2)I \end{bmatrix} \quad (48)$$

And following equalities hold:

$$\begin{bmatrix} w(x)\frac{\partial g_1}{\partial x_1} - \sum_{r=1}^m L_{1r}B_{1r} & \cdots & w(x)\frac{\partial g_1}{\partial x_n} - \sum_{r=1}^m L_{1r}B_{nr} \\ \vdots & \ddots & \vdots \\ w(x)\frac{\partial g_m}{\partial x_1} - \sum_{r=1}^m L_{mr}B_{1r} & \cdots & w(x)\frac{\partial g_m}{\partial x_n} - \sum_{r=1}^m L_{mr}B_{nr} \end{bmatrix} = 0 \quad (49)$$

Where

$$\psi_1 = GAP + PA^T G^T - \gamma GB_2 B_2^T G^T - \sum_{j \in J} \frac{\partial P}{\partial x_j}(A_j Z) + \varepsilon_2 I$$

$P(\tilde{x})$ and $L(x)$ are $N \times N$ and $m \times m$ symmetric polynomial matrices respectively.

**Proof**: Again, following the same procedure as in theorem 1, if (46) is satisfied, maintenance of sliding mode is guaranteed. The equivalent control law is then achieved:

$$u_{eq} = (I + \varphi_0(x,t))^{-1}(K(x) - \varphi_2(x,t) - (M(x)B(x))^{-1}M(x)B^\perp \varphi_3(x,t))$$

And sliding mode dynamics is described as:

$$\dot{x} = f(x) + B(x)k(x) + \varphi_{eq}(t,x)$$

Now in order to stabilize the closed loop system and design the state feedback $k(x)$, SOS based $H\infty$ technique is utilized [10]. Consider the system with artificial penalty variable $z = [z_1 \ z_2]$ as follows:

$$\begin{bmatrix} \dot{x} \\ z_1 \\ z_2 \end{bmatrix} = \begin{bmatrix} A(x) & B_1(x) & B_2(x) \\ C_1(x) & 0 & 0 \\ 0 & 0 & I \end{bmatrix} \begin{bmatrix} Z(x) \\ \varphi_{eq} \\ u \end{bmatrix} \quad (50)$$

Where $Z(x)$ is a monomial vector satisfying assumption $Z(x) = 0$ if $x = 0$. The objective here is to design a state feedback $k(x)$ for the system above with penalty variable $z$ such that the $L_2$-gain of the transfer matrix $T_{z\varphi_4}$ is minimized, optimizing the performance index $\gamma^2$:

$$T_{z\varphi_4} = \frac{\int_0^T \|z\|^2}{\int_0^T \|\varphi_4\|^2} \leq \gamma^2 \quad (51)$$

Influenced by [10, 20] and similar to the proof of the theorem 3 the proof is completed. ∎

*Remark 4*: For such an $M(x)$, $w(x)$ plays an important role in existence of $g(x)$ since it extends the transformation polynomial matrix to the rational case. This point has been illustrated by example 2 in section VI.

## VI. ILLUSTRATIVE EXAMPLES

In this section, some examples are provided to show the applicability and flexibility of the method developed in this paper. It should be noted that anywhere needed, the SOS programs are solved by means of SOSTOOLS.

*Example 1*: In this example, two approaches are proposed to show that various models can be formatted to fit the method's requirement. Consider the nonlinear time varying uncertain system

$$\begin{cases} \dot{x}_1 = -x_1^3 + x_2 - x_3 e^{-2t} + x_3^{2/3} u(t) + u(t) \\ \dot{x}_2 = -x_1 - x_2 + 0.3x_1 \cos(x_1) + px_3 + 0.01t \\ \dot{x}_3 = -x_3 + 0.1x_2 \sin(x_1) + 0.05 \sin(\pi t) \end{cases}$$

in which $p$ is an uncertain value, bounded by $\bar{p} = 0.005$ and $\underline{p} = 0.003$ as $\underline{p} \leq p \leq \bar{p}$. In order to transform the system equations to the form of (3), non-polynomial and time-varying terms are considered as perturbations. Therefore the polynomial system with uncertainty is obtained as

$$f(x) = \begin{bmatrix} -x_1^3 + x_2 \\ -x_1 - x_2 + px_3 \\ -x_3 \end{bmatrix}, B(x) = \begin{bmatrix} 1 \\ 0 \\ 0 \end{bmatrix}, \varphi_0(t,x) = \begin{bmatrix} x_3^{2/3} \\ 0 \\ 0 \end{bmatrix},$$

$$\xi(t,x) = \begin{bmatrix} -x_3 e^{-2t} \\ 0.3x_1 \cos(x_1) + p_{nominal} x_3 + \Delta p x_3 + 0.01t \\ 0.1x_2 \sin(x_1) + 0.05\sin(\pi t) \end{bmatrix}$$

Where $\Delta p_1$ is the variation of $p_1$ around its nominal value. It is also possible to increase the bounds of uncertainty in this method. To this end, we can use the recasting technique for elementary functions. The slack variables are defined by

$$x_4 = \cos(x_1)$$
$$x_5 = \sin(x_1)$$
$$x_6 = t$$
$$x_7 = x_3^{1/3}$$
$$x_8 = x_3^{-1/3}$$

The equivalent system using these variables is then:

$$f(x) = \begin{bmatrix} -x_1^3 + x_2 \\ -x_1 - x_2 + 0.3x_1 x_4 + 0.01x_6 \\ -x_3 + 0.1x_2 x_5 \\ -x_5(-x_1^3 + x_2) \\ x_4(-x_1^3 + x_2) \\ 1 \\ x_8^2(-x_3 + 0.1x_2 x_5)/3 \\ -x_8^4(-x_3 + 0.1x_2 x_5)/3 \end{bmatrix}, B(x) = \begin{bmatrix} x_7^2 + 1 \\ 0 \\ 0 \\ \vdots \\ 0 \end{bmatrix}$$

$, \varphi_0(t,x) = 0,$
$\xi(t,x) = [-x_3 e^{-2t} \quad px_3 \quad 0.05\sin(\pi t) \quad 0 \quad \cdots \quad 0]^T$

The constraints are

$x_4^2 + x_5^2 = 1$
$x_7^3 - x_3 = 0$
$x_7 x_8 - 1 = 0$
$x_6 > 0$

*Example 2*: This example shows the applicability of the proposed method. In this example matched perturbation Is considered.

$$f(x) = \begin{bmatrix} -x_1 + x_2 \\ x_1^2 - 2x_2 x_1^2 - x_2^3 - x_2 \end{bmatrix}, B(x) = \begin{bmatrix} 0 \\ 1 \end{bmatrix},$$
$\varphi_0(t,x) = 0.1\sin^3(x_1)$

$\varphi_1(x,t) = 0.1x_1 cos^2(x_2) + 0.1 \sin(\pi t)$

The bounds are $\beta_1 = 0.1$ and $\rho(x,t) = 0.1|x_1| + 0.1$. let the initial state of the system be $x(t_0) = [0.2 \quad 0.5]^T$. SOS programming of theorem 2 can be used to show that the closed loop system is asymptotically stable. In this example, matrix $Q(x_1)$ is considered to be degree 1. Design parameters are: $q(x_1) = 1, \varepsilon_1 = 0.1, \quad \varepsilon_2 = 0.01$ and $Z(x) = [x_1, x_2]^T$. The following results are obtained.

$Q(x_1) = \begin{bmatrix} Q_{11}(x_1) & Q_{12}(x_1) \\ Q_{12}(x_1) & Q_{22}(x_1) \end{bmatrix}$
$Q_{11}(x_1) = +0.88725$
$Q_{12}(x_1) = 0$
$Q_{22}(x_1) = 0.7174$
$N(x) = [N_1(x) \quad N_2(x)]$
$N_1(x) = -0.022447x_1 - 0.13418x_2 - 0.39933$
$N_2(x) = -0.13418x_1 - 0.17093x_2 + 0.060698$
$\rho(x,t) = \|N(x)Q^{-1}(x_1)Z(x)\|_2 + 1 + |x_1|$
$g(x) = 0.982x_2$

The corresponding controller can be expressed as

$u(t) = \begin{cases} N(x)Q^{-1}(x_1)Z(x) - \rho(x,t)sign(s)s(x,t) \neq 0 \\ N(x)Q^{-1}(x_1)Z(x) \quad s(x,t) = 0 \end{cases}$

The state trajectory of the closed loop system and the control signal are illustrated in Fig.1 which shows that closed loop is asymptotically stable.

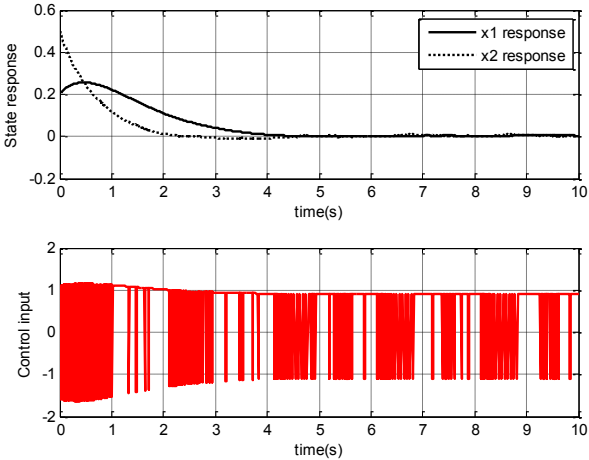

Fig. 1. Closed-loop response and control input signal for example3:matched perturbation.

*Example 3*: This example investigates applicability of our method to systems with both matched and unmatched perturbations. Consider the case in which an unmatched perturbation is added to the system of example 2.

$\varphi_2(x,t) = 0.1 \sin(x_2) + x_1 \sin(\pi t), B^\perp = [1 \quad 0]^T$

Here $P(x_1)$, degree 2 matrix, has been designed with $w(x) = 1$. Theorem 4 results in:

$P(x_1) = \begin{bmatrix} P_{11}(x_1) & P_{12}(x_1) \\ P_{12}(x_1) & P_{22}(x_1) \end{bmatrix}$
$P_{11}(x_1) = -2.9402x_1 + 0.40762$
$P_{12}(x_1) = -0.629x_1 + 0.1635$
$P_{22}(x_1) = 1.6709x_1^2 - 1.7584x_1 + 1.6068$
$\gamma = 0.25148$
$g(x) = 0.91184x_2$

and the corresponding controller can be obtained by (45). Closed loop response for the system in example 3 with unmatched perturbation and the corresponding control signal are illustrated in Fig. 2 which shows asymptotic stability of the origin.

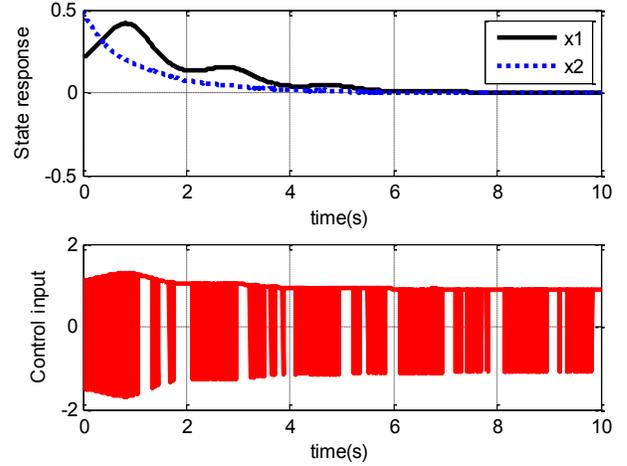

**Fig. 2.** Closed-loop response and control input signal for example4:unmatched perturbation.

## VII. CASE STUDIES

### A. Glucose- insulin interaction in blood system

The proposed method is now applied to Glucose and insulin interaction in blood system. One of the renown models for this, known as Bergman's minimal model as follows [21]

$\begin{cases} \dot{G}(t) = -p_1(G(t) - G_b) - X(t)G(t) + D(t) \\ \dot{X}(t) = -p_2X(t) + p_3(I(t) - I_b) \\ \dot{I}(t) = -n(I(t) - I_b) + \gamma(G(t) - h)^+ t + u(t) \end{cases}$ (54)

Where $t = 0$ is the time that glucose is injected to vein, '+' sign is the positive reflection to glucose intake, $G(t)$ is the glucose concentration in blood plasma $(mg/dl), X(t)$ is the insulin's effect on the net glucose disappearance which is referred to as the remote insulin concentration $(1/min)$, $G_b$ is the basal pre-injection level of glucose $(mg/dl)$, $I(t)$ is the insulin concentration in plasma at time $t$ $(\mu U/ml)$, $I_b$ is the basal pre-injection level of insulin $(\mu U/dl)$ and $D(t)$ shows the rate at which glucose is absorbed into the blood from intestine. Since normal insulin regulatory system does not exist in diabetic patients, this glucose absorption is considered a disturbance for the system dynamics, and it can be modeled by a decaying exponential function in which $p_1$ is the insulin-dependent rate constant of glucose consumption in muscles and liver ($1/min$), $p_2$ is the rate for decrease in tissue glucose uptake ability ($1/min$), $p_3$ is the insulin-dependent increase in glucose uptake ability in tissue per unit of insulin concentration above the basal level (($\mu U/ml$)/$min2$), $n$ is the first order decay rate for insulin in blood ($1/min$), $h$ is the threshold value of glucose above which the pancreatic $\beta$ −cells release insulin (mg/dl), and $\gamma$ is the rate of pancreatic $\beta$-cells' release of insulin after glucose injection with glucose concentration above the threshold (($\mu U/ml$)/$min2$/($mg/dl$)). The time-varying term is considered as disturbance and the equilibrium point is moved to the origin by a simple state transformation:

$$x_1(t) = G(t) - G_b$$
$$x_2(t) = X(t)$$
$$x_3(t) = I(t) - I_b$$

Regarding [22], system parameters are considered with perturbation. In order to take parametric uncertainty of system into account, uncertainty bounding set $\theta$ is defined as

$$\theta = \{(p_2, p_3, n, \gamma, h) |\ \underline{p_2} \le p_2 \le \overline{p_2}, \underline{p_3} \le p_3 \le \overline{p_3}, \underline{n} \le n \le \overline{n}, \underline{h} \le h \le \overline{h}, \underline{\gamma} \le \gamma \le \overline{\gamma}\}$$

With these considerations, the system dynamics can be represented by the following set of equations.

$$\dot{x}_1(t) = -p_1 x_1 - x_1 x_2 + G_b x_2 + D(t)$$
$$\dot{x}_2(t) = -p_2 x_2 + p_3 x_3$$
$$\dot{x}_3(t) = -n x_3 + u(t) + \gamma(s + 0.5)(x_1 + G_b - h)t$$
$$\dot{t} = 1$$

Where the following equality and inequality constraints are satisfied.

$$(s + 0.5)(s - 0.5) = 0$$
$$s(x_1 + G_b - h) \ge 0$$
$$\alpha_1 = -s(x_1 + G_b - h)$$
$$\alpha_2 = (p_2 - \underline{p_2})(p_2 - \overline{p_2})$$
$$\alpha_3 = (p_3 - \underline{p_3})(p_3 - \overline{p_3})$$
$$\alpha_4 = (n - \underline{n})(n - \overline{n})$$
$$\alpha_5 = (\gamma - \underline{\gamma})(\gamma - \overline{\gamma})$$
$$\alpha_6 = (h - \underline{h})(h - \overline{h})$$

Incorporating these equality and inequality constraints into SOS program of theorem 4, sliding mode manifold parameters and control parameters are obtained as:

$$g(x) = 0.95378 x_3(t)$$
$$L = \begin{bmatrix} 0.95378 & 0 & 0 \\ 0 & 0.95378 & 0 \\ 0 & 0 & 0.95378 \end{bmatrix}$$

Moreover, in order to reduce chattering effect, a linear low pass filter is applied to smooth the discontinuous control function.

$$u(t) = \begin{cases} -\gamma B_2^T G^T P^{-1}(\tilde{x}) Z(x) - \rho(x,t) \dfrac{\left(L(x) B^T(x) w^{-1}(x) B\right)^T s}{\left\|\left(L(x) B^T(x) w^{-1}(x) B\right)^T s\right\|} & if\ \|s\| \ge \alpha \\ -\gamma B_2^T G^T P^{-1}(\tilde{x}) Z(x) & if\ \|s\| < \alpha \end{cases}$$

where $\alpha = 0.05$. Applying aforementioned control to the system we obtained the state trajectories plotted in Fig.3 and Fig.4. Moreover, the control function is depicted in Fig.5. As seen, Fig.3 and Fig.4 show Glucose and Insulin response for three patients which indicate asymptotic stability of the equilibrium point. The current paper proposed ISM is designed systematically which is indeed a main advantage of this paper's method.

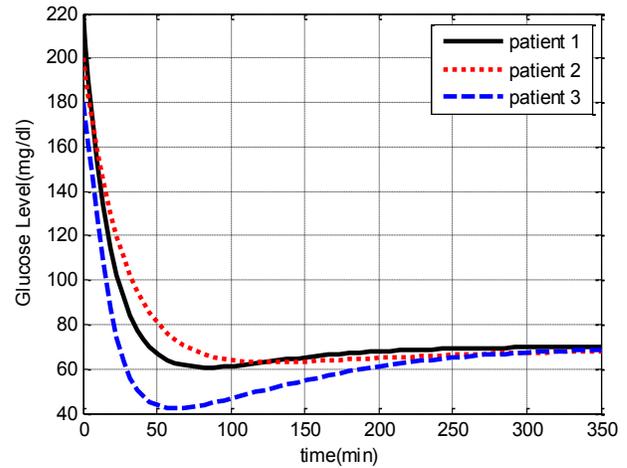

**Fig. 3.** Closed-loop glucose regulatory system using the proposed method.

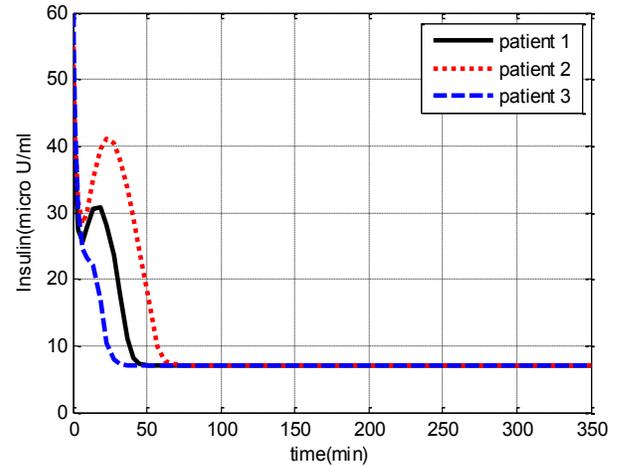

**Fig. 4.** Closed-loop insulin profile..

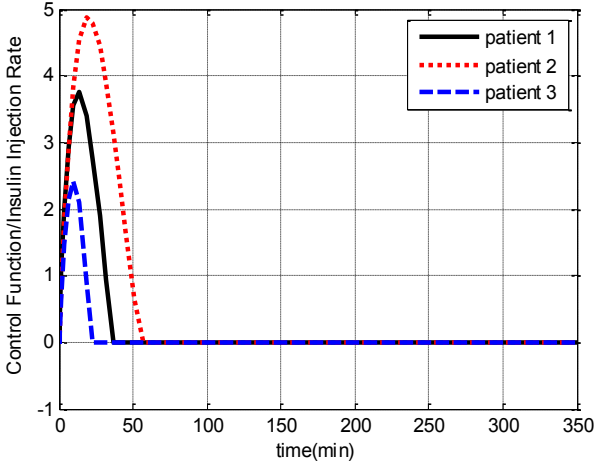

**Fig. 5.** Control function.

### B. Closed-loop steering of Unicycle

In this section, the proposed method is applied to the unicycle's dynamics. This example is intentionally given to compare effectiveness of this paper's framework with that of [4, 23, 24].

The unicycle's dynamics with matched and unmatched perturbations is given below

$$\begin{cases} \dot{x}_1 = (u + \varphi_{11})cosx_3 - \varphi_2 sinx_3 \\ \dot{x}_2 = (u + \varphi_{11})sinx_3 + \varphi_2 cosx_3 \\ \dot{x}_3 = w + \varphi_{12} \end{cases}$$

In the above, $\varphi_1 = [\varphi_{11}, \varphi_{12}]^T$ and $\varphi_2$ respectively represent matched and unmatched uncertainties, and are given by

$$\varphi_{11} = 1.2\sin(5t), \varphi_{12} = 0.4\sin(5t), \varphi_2 = 0.8\sin(t)$$

Now, in order to stabilize the unicycle's dynamics without alteration in coordination, dynamic control law must be exploited. Regarding this objective, new variables are defined as

$$\begin{cases} z_1 = x_1 \\ z_2 = x_2 \\ z_3 = \dot{x}_1 \\ z_4 = \dot{x}_2 \\ \dot{\xi} = A_1 \end{cases}$$

in which $\xi \in \mathbb{R}$ is generated by a nonlinear function $A_1$. By this definition, system's nominal dynamics can be described as in the following.

$$\begin{cases} \dot{z}_1 = u_1 \cos(x_3) \\ \dot{z}_2 = u_1 \sin(x_3) \\ \dot{z}_3 = \dot{u}_1 \cos(x_3) - u_1 u_2 \sin(x_3) \\ \dot{z}_4 = \dot{u}_1 \sin(x_3) + u_1 u_2 \cos(x_3) \\ \dot{\xi} = A_1 \end{cases}$$

Using recasting procedure similar to the previous approach, the problem can be tested by a SOS program. We consider $u_1 = \xi$ and $u_2 = A_2(z_1, z_2, z_3, z_4, \xi)$ to simplify the SOS programming. The following results are obtained.

$$g_1(x) = x_3, g_2(x) = x_1 cos(x_3) + x_2 sin(x_3)$$
$$A_1 = (0.9152x_1 + 0.785z_3) cos(x_3)$$
$$\qquad - (0.9152x_2 + 0.785z_4) sin(x_3)$$
$$A_2 = \begin{bmatrix} (0.9152x_1 + 0.785z_3) sin(x_3) \\ -(0.9152x_2 + 0.785z_4) cos(x_3) \end{bmatrix} / \xi$$

The solution given in [4] is indeed a special case of this programming. This approach introduces a set of systematically obtained solutions for this problem with one of them given in [4]. In addition, the approach in [23] is based on feedback linearization while nonlinear Lyapunov function technique underpins this paper's approach. Fig.6 compares the result of this paper with that of [23] in terms of closed loop response.

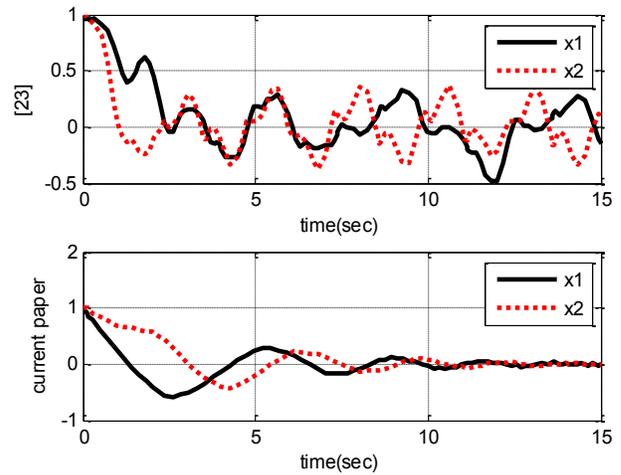

**Fig. 6.** Closed-loop response unicycle.

Fig.6 suggests that the set regulation problem's response has a much better performance when our GISM controller is hired. It causes a relatively fast convergence to origin while the method of [23] results in many fluctuations in states, and needs more time for regulation. Since method have been proposed in [23] is sensitive to perturbation and its controller have been designed to stabilize only the system without any perturbation. Furthermore, the current paper ISM is designed systematically which is in fact an important benefit of this paper's method.

The current paper presented ISM can be extended to address this problem in terms of polar coordinates. In order to solve set point regulation problem for the dynamics, the position of the vehicle in terms of its polar coordinates is used [24]. By introducing these new variables, the system's equations become

$$\begin{cases} \dot{e} = -(u + \varphi_{11})cos\alpha + \varphi_2 sin\alpha \\ \dot{\alpha} = -(w + \varphi_{12}) + (u + \varphi_{11})(sin\alpha/e) \\ \dot{\theta} = (u + \varphi_{11})(sin\alpha/e) + \varphi_2 e cos\alpha \end{cases}$$

According to remark 1, $x_4 = sinx_3$ and $x_5 = cosx_3$ are

defined as slack variables. Hence, $\bar{x}_1 = [x_1, x_2, x_3]$, $\bar{x}_2 = [x_4, x_5]$ and $G_1(\bar{x}_1, \bar{x}_2) = x_4^2 + x_5^2 - 1$ are used to recast non-polynomial system in a polynomial one. Using SOS program theorem 3 the results are obtained as:

$$g(x) = \begin{bmatrix} -0.5x_1^2 \cos(x_2) + x_3 \sin(x_2) - \cos(x_2) \\ -x_2 \end{bmatrix}$$

$$K = [K_1 \quad K_2]^T$$
$$K_1 = 1.482\cos(x_2)$$
$$K_2 = 1.4693 x_2 + 1.482 \sin(x_2) \cos(x_2) + 0.482 \frac{x_3}{x_2} \sin(x_2) \cos(x_2)$$

Fig.7 shows the unicycle's closed loop response for the current paper presented ISM and the proposed method of [24]. As seen, the closed loop response resulted from the method of [24] has a poor disturbance rejection when disturbances are introduced. On the contrary, the proposed ISM shows significant improvement in closed loop response as states uniformly converge to zero. Moreover, all parameters of the sliding surface and control are obtained algorithmically which is indeed an important advantage of this paper's method.

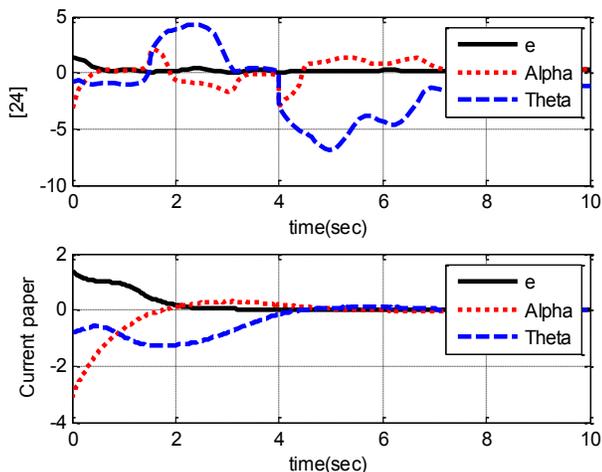

**Fig. 7.** Closed-loop response unicycle using polar coordination.

## VIII. Conclusion

A new method for design of nonlinear integral sliding mode control based on the Sum Of Squares has been developed in this paper. Nonlinear systems with matched and unmatched perturbations have been discussed separately. Several examples were presented to verify applicability of the proposed method. Some examples are also included to show that various models can be formatted to fit to the method's requirements. Benefits of this approach can be summarized as 1) to provide a systematic approach for designing a sliding mode controller, and 2) existence of efficient numerical methods for solving the problem. For further improvement one can extend the theorems in order to stability achieved by means of finite time stability instead of asymptotically stability.